\documentclass[pre,superscriptaddress,twocolumn,nofootinbib]{revtex4}
\usepackage{graphicx} 
\usepackage{amsmath}

\newcommand{\br}{\mathbf{r}}
\newcommand{\bn}{\begin{equation}}
\newcommand{\ee}{\end{equation}}
\newcommand{\half}{\frac{1}{2}}
\newcommand{\diff}{\mathrm{d}}

\newcommand{\vdW}{{\mbox{\scriptsize vdW-DF}}}
\newcommand{\GGA}{{\mbox{\scriptsize GGA}}}
\newcommand{\LDA}{{\mbox{\scriptsize LDA}}}

\newcommand{\tot}{{\mbox{\scriptsize tot}}}
\newcommand{\cut}{{\mbox{\scriptsize cut}}}
\newcommand{\nl}{{\mbox{\scriptsize nl}}}

\newcommand{\reff}{{\mbox{\scriptsize ref}}}

\newcommand{\system}{{\mbox{\scriptsize system}}}
\newcommand{\adenine}{{\mbox{\scriptsize adenine}}}
\newcommand{\graphite}{{\mbox{\scriptsize graphite}}}
\newcommand{\overlayersheet}{{\mbox{\scriptsize o-s}}}
\newcommand{\sheetchain}{{\mbox{\scriptsize s-r}}}
\newcommand{\chainmolecule}{{\mbox{\scriptsize r-m}}}
\newcommand{\Delto}{\Delta^{\!0}}

\newcommand{\chalmersTF}{Department of Applied Physics, Chalmers University of Technology,
SE-41296 G\"{o}teborg, Sweden}
\newcommand{\chalmersMC}{Department of Microtechnology and Nanoscience, MC2,
Chalmers University of Technology,
SE-41296 G\"{o}teborg, Sweden}

\newcommand{\rutgers}{Department of Physics and Astronomy, Rutgers University,
Piscataway, NJ 08854-8019, USA}
\newcommand{\oakridge}{Materials Science and Technology Division, 
Oak Ridge National Laboratory, Oak Ridge, TN 37831-6114, USA}
\begin{document}

\title{A van der Waals density functional study of adenine on graphene: \\
Single molecular adsorption and overlayer binding}

\author{Kristian Berland}\affiliation{\chalmersMC}
\author{Svetla D. Chakarova-K{\"a}ck}\affiliation{\chalmersTF}
\author{Valentino R. Cooper}\affiliation{\rutgers}\affiliation{\oakridge}
\author{David C. Langreth}\affiliation{\rutgers} 
\author{Elsebeth Schr\"oder}\affiliation{\chalmersMC} 

\date{September 28, 2010}

\begin{abstract}
The adsorption of an adenine molecule on graphene is studied using a first-principles van der
Waals functional (vdW-DF) [Dion et al., Phys.\ Rev.\ Lett.\ \textbf{92}, 246401 (2004)]. 
The cohesive
energy of an ordered adenine overlayer is also estimated. For the adsorption of a single molecule, we determine the
optimal binding configuration and adsorption energy by translating and rotating the molecule. The
adsorption energy for a single molecule of adenine is found to be 711 meV, which is close to the
calculated adsorption energy of the similar-sized naphthalene. Based on the
single molecular binding configuration, we estimate the cohesive energy of a two-dimensional ordered overlayer.
We find a significantly stronger binding energy for the ordered overlayer than for single-molecule
adsorption. 
\end{abstract}

\maketitle

\section{Introduction}

Physisorption of small biomolecules on inert surfaces acts as 
a laboratory of molecular interactions and is an excellent 
starting point for addressing molecular recognition and self-organization processes.
By studying these systems we gain insight into the delicate balance 
between intermolecular forces that contribute to supramolecular binding,
for example, the unique identification 
of antigens and their binding sites in our biochemistry.  

It is natural to begin an investigation 
of molecular interactions and organization by focusing on key building blocks,
like the nucleic acids or the amino acids.  Three-dimensional 
biopolymer systems, like the DNA double helix or proteins, permit folding 
of an extreme complexity that limits  
direct access to most of the structure,
and direct experimental probing of the atomic-scale organization is difficult. 
Theoretical modeling must therefore map out interactions and organization not 
only between primary but also secondary and higher order structure 
in the absence of any calibration with experiments.  In contrast, trapping 
nucleobases on inert surfaces not only  simplifies the possibilities
for structural reorganizations but also makes the molecular interactions accessible to 
direct characterization through advanced atomic-scale experiments.

Adenine is one of the nucleobases of DNA. The molecule has been investigated in numerous 
sophisticated surface experiments that 
have characterized both the physi\-sorp\-tion and self-organization or overlayer formation 
on inert substrates, ranging from the insulating MoS$_2$ \cite{sowerby2000} over the semi-metallic 
graphite \cite{srinivasan1993,tao1994,freundthesis,freund,sowerby2000} 
to surfaces of noble metals like Cu and Ag \cite{chen2002,feyer2009}.
The experimental characterization includes thermal desorption spectroscopy (TDS) 
\cite{freundthesis} for a direct measurement of the physisorption energy, and 
scanning tunneling microscopy \cite{freund,srinivasan1993,tao1994,sowerby2000}, 
atomic force microscopy \cite{tao1994}, and low-energy electron diffraction \cite{freund} for 
explicit identification of the adenine assembly into regular overlayers.

Here we apply density functional theory calculations (DFT) with a fully nonlocal
density-functional method, vdW-DF \cite{vdW1,vdW1sc}, that provides a 
first-principle account of dispersive or van der Waals (vdW) forces. The 
method permits a parameter-free description of a broad spectrum of sparse matter 
\cite{Review:vdW}: materials which have regions with 
voids in the electron distributions, such as molecular systems. 
Unlike the semi-empirical DFT-D methods, it does
not involve an arbitrariness in the construction of a damping function.
The vdW-DF method has previously 
been used to describe binding in a large range of material systems, for example 
dimers of benzene \cite{vdW1sc,PAH:dimer}, nucleobases \cite{jacs,DNA}, polymers \cite{Jesper:vdW}, 
nanotubes \cite{Nanotube:vdW}, simple oxides \cite{V2O5}, 
and molecular-crystals \cite{KB:Molcrys,berlandX} systems. It has furthermore 
been used to characterize the physisorption of organic molecules on coinage metals
(Au, Ag, Cu) \cite{vdWDF:Japan,PTCDA:coinage,KB:benzCu,Wellendorff2010,Mura2010}, on
MoS$_2$ \cite{MOS2}, on Si \cite{Karen:BenzeneSilicon}, on alumina \cite{Svetla:phenol}, and on
graphene \cite{Svetla:phenol,Svetla:BenzeneGraphite}. In short, it is a versatile method. 

By comparing directly with experimental measurements the vdW-DF method 
has been documented to provide good results
for benzene and naphthalene on graphene \cite{zacharia,Svetla:BenzeneGraphite}. 
This suggests that vdW-DF
should provide a good description also of the adenine physisorption.
The simple form of the non-local correlations in vdW-DF allows for efficient calculations 
of the adenine physisorption and of the mutual adenine interactions 
leading to the formation of a two-dimensional overlayer crystal. These
results can be directly compared with experiments. 
For an adenine molecule on a graphite sheet we here find an adsorption energy 
of 0.711 eV at an equilibrium separation of 3.5 {\AA}, whereas the adenine molecules
adsorb with the energy 1006 meV per molecule in the overlayer crystal in very good agreement 
with experiment \cite{freundthesis}. 

In the following, section II describes how we use the vdW-DF method for the present
adsorbate system. In section III we discuss the framework of the method 
in relation to some other methods used for
this system, and in section IV we present and discuss our results for the configuration and
binding of the single adenine adsorbate and an estimate of the binding energy of adenine in the
overlayer. Secction V contains our conclusions.
 
\section{Computational method}

We use the first-principles vdW-DF method within DFT \cite{vdW1}, calculating 
the vdW-DF energies in a post-GGA procedure similar to previous studies 
\cite{Svetla:phenol,Svetla:BenzeneGraphite,Eleni:vdW,PAH:dimer,KB:benzCu,V2O5}. 
We calculate the vdW-DF total energy, $E^\vdW$, for a number of positions of the 
adenine molecule above a graphite sheet, as
described in Section IV. 
For each position, a self-consistent GGA (sc-GGA) 
calculation is carried out, from which the GGA-based total 
energy $E^\GGA_\tot$ and the sc-GGA electron density $n$ are obtained.
In a post-processing phase, we use $n$ to evaluate long-range correlation contributions
that arise from the vdW interactions, $E_c^\nl$ and then in a systematic way combine the 
sc-GGA results and the nonlocal results to obtain $E^\vdW$. This procedure is 
described below and is further detailed in several other publications 
\cite{Svetla:BenzeneGraphite,Eleni:vdW,PAH:dimer}.

For the sc-GGA calculations we utilize the plane-wave code \textsc{dacapo} \cite{DACAPO}
with the PBE exchange-correlation \cite{PBE}. We use ultrasoft pseudopotentials, a 
$2 \times 2 \times 1$ sampling of the Brillouin zone in the
Monkhorst-Pack scheme, a wavefunction energy cutoff at 500 eV, and a fast Fourier transform (FFT)
grid with a maximum of 0.15 {\AA} between nearest-neighbor gridpoints. 

The correlation part of the energy in $E^\vdW$ can be split into a nearly-local part
$E^0_c$ and a part that includes the most nonlocal interactions $E_c^\nl$,
\bn
E_c=E^0_c+E_c^\nl.
\ee 
The nearly-local part is approximated by the correlation of the local density 
approximation (LDA) $E_c^\LDA$, and 
the nonlocal correlation functional is given by the integral
\bn
E^\nl_c[n]= \half \int \diff \br \, \diff \br' \, n(\br) \phi(\br,\br') n(\br')\,.
\label{eq:gg}
\ee
Ref.~\onlinecite{vdW1} contains the explicit form of the kernel $\phi$.
The vdW-DF total energy can thus be written as the sum $E^\vdW=E^0+E^\nl_c$  where
the  $E^0$ term includes kinetic and electrostatic terms in addition to GGA exchange 
and LDA correlation:
\bn
E^{0}=E^\GGA_\tot-\left(E^\GGA_c-E^\LDA_c\right)\,. 
\label{Master1}
\ee
This splits off the nonlocal part of the calculation, that needs a slightly different
treatment than the nearly-local\footnote{Local in correlation.} term $E^0$.

As in recent applications \cite{Eleni:vdW,PAH:dimer,Jesper:vdW,KB:Molcrys}
of vdW-DF we use the revPBE \cite{revPBE} GGA exchange in the post-processing 
phase. 
From the total energy of the sc-GGA calculations we therefore subtract the 
PBE exchange energy and instead add the revPBE exchange energy. The revPBE exchange
energy is calculated 
from the charge density $n$ that is provided by the sc-GGA calculations. The revPBE 
exchange functional is known to be overly repulsive in the binding region \cite{vdW1}.
For a range of systems, vdW-DF with revPBE exchange
provides good values for the binding energy and quite good, 
but consistently overestimated binding separations. The development of an 
exchange-functional companion to the non-local correlation of vdW-DF is an active 
research field showing 
promising results \cite{KlimesComment,CooperX,LangrethX,V2O5,londero2}.

\begin{figure}[tb]
\centering
\includegraphics[width=5cm]{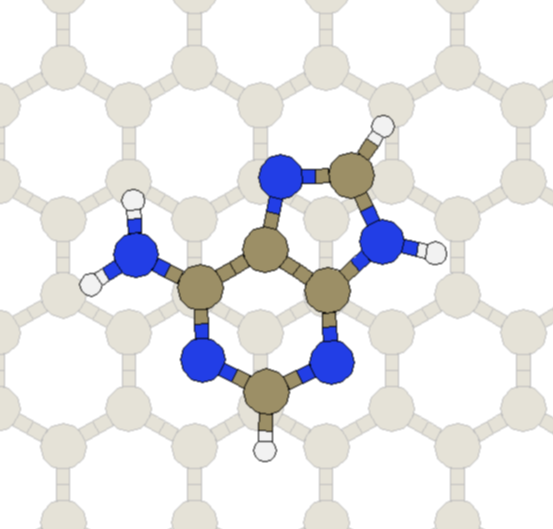}
\caption{(Color online) The molecular configuration of adenine on 
graphite as determined in our vdW-DF study.
Dark (blue) circles are N atoms, medium gray (brown) circles are C atoms, and the small white
circles are H atoms. The light-colored background illustrates the underlying graphite sheet.}
\label{fig:isolated}
\end{figure}

The adsorption energy is given by the difference between $E^\vdW$ for the 
optimal configuration and for a reference system 
corresponding to isolated fragments (molecules). 
Since intra-molecular and intra-sheet contributions dominate $E^\vdW$ we must 
use the same numerical approximations in the two calculations. 
To conveniently cancel parameter-sensitive contributions, the reference 
calculation for the sc-GGA part is done in a manner different from the non-local 
correlation part. The difference in total energy $E^\vdW$
of the adsorbate system compared to that of separated fragments, the 
cohesive energy $E$, is thus the sum of two terms
\bn
E = \Delto E^0+ \Delta^\nl E^\nl_c 
\ee
where the $\Delto$ and $\Delta^\nl$ denote the use of two different sets of 
reference calculations. At the optimal position of the adenine on graphite 
(Fig.\ \ref{fig:isolated}) the binding energy $E_b$ is 
found (with positive sign convention for binding).

\subsection{Reference calculations for the sc-GGA part}

For the calculations of the single adenine molecule adsorbed on the graphite sheet
the reference sc-GGA calculation uses the same 
unit cell as the adsorbate system, but  
in a configuration where the adenine molecule is lifted 9 {\AA} away from the 
graphene sheet:
\bn
\Delto E^0 = E^0 -E^0_\reff
\ee 
where $E^0_\reff$ is the reference calculation.
This distance of 9 {\AA} is fully sufficient for the GGA calculations which only include 
interactions acting at a much smaller distance. 
By using the same unit 
cell for the reference calculation as for the full calculation, we cancel a small spurious 
contribution from the regions of very low electron density in the sc-GGA 
calculation \cite{PAH:dimer,V2O5,Offset,londero2}.

\begin{figure*}[tb]
\centering
\includegraphics[width=15cm]{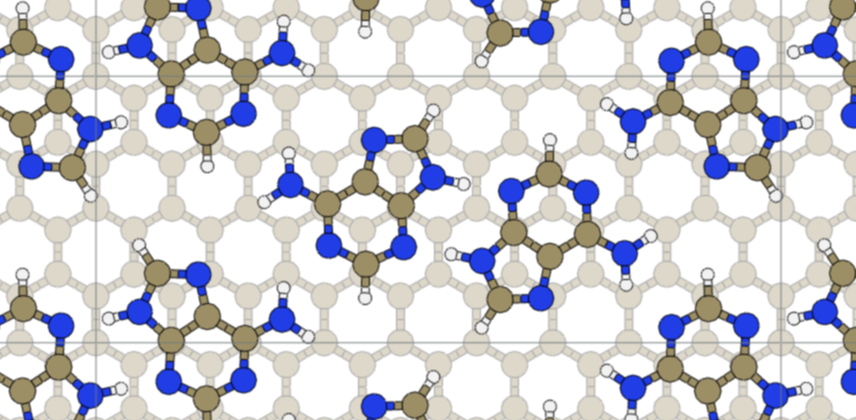}
\caption{(Color online) A two-dimensional adenine crystal on the surface of graphite:
The configuration used in our estimate of the binding energy. There are
four molecules in the rectangular unit cell. Same color coding of atoms as
used in Fig.\ \protect\ref{fig:isolated}.}
\label{fig:crystal}
\end{figure*}

\begin{figure*}[tb]
\centering
\includegraphics[width=15cm]{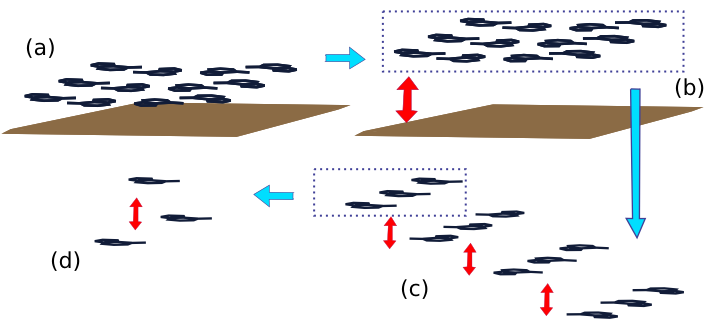}
\caption{(Color online) Sketch of the procedure for calculating the contributions to $\Delto E^0$
from the adsorption of the adenine overlayer. (a): The system of adsorbed adenine molecules
(black) in a two-dimensional crystal on the graphite surface (medium gray/brown).
(b): Lifting off the sheet of adenine molecules in order to calculated  $\Delto E^0_\overlayersheet$.
(c): Creating ribbons of adenine molecules (each of the four molecules in a 
unit cell participate in a different ribbon) by moving the molecules apart in one 
direction, to calculate $\Delto E^0_\sheetchain$.
(d): Each ribbon is taken apart, yielding the energy $\Delto E^0_\chainmolecule$. }
\label{fig:assembly}
\end{figure*}

In the calculations for the overlayer of adenine on  graphite the contributions
to $\Delto E^0$ are obtained in three steps, as indicated in Figure \ref{fig:assembly}.
In each step, the reference calculation uses the same unit cell as the full calculation.
First, the adenine overlayer crystal (o), shown in Figure \ref{fig:crystal} and 
sketched in Figure \ref{fig:assembly}.a,  is lifted off 
as an intact sheet (s) from the graphite sheet to a distance 9 {\AA} above the graphite sheet
(Fig.\ \ref{fig:assembly}.b).
We denote the energy cost per molecule of this process by $\Delto E^0_\overlayersheet$.
Then the sheet of the adenine crystal is split into ribbons (r) of width one adenine molecule
(Fig.\ \ref{fig:assembly}.c), with energy cost $\Delto E^0_\sheetchain$ per molecule,
and finally the ribbons are disassembled into individual molecules (m), 
$\Delto E^0_\chainmolecule$ (Fig.\ \ref{fig:assembly}.d).
In total, the contribution to the overlayer cohesion
energy per adenine molecule is
\begin{equation}
\Delto E^0 = \Delto E^0_\overlayersheet +\Delto E^0_\sheetchain+\Delto E^0_\chainmolecule \,.
\end{equation}
The quantity $\Delto E^0$ could also have been obtained by simply taking 
all five fragments (graphite sheet and four adenine molecules) far apart within 
the unit cell, but this would require an unreasonably large unit cell, both 
for the GGA reference calculations and all other GGA calculations of the molecule 
overlayer on graphite. 

\subsection{Reference calculations for the nonlocal part}

Our implementation of (\ref{eq:gg}) is sensitive to the choice of grid on which the
charge density is described. To avoid adverse effects of this sensitivity we use
a charge-density grid with a volume per grid-point smaller than
(0.15 {\AA})$^3$ in all our \textsc{dacapo} calculations. 
Further, for every adsorbed adenine
configuration we carry out a separate reference calculation of the isolated molecule
where the molecule is locally placed in the same position relative to the charge density grid.  
A similar reference calculation for 
the graphite sheet is carried out once, since this sheet is kept fixed. 
The contribution of the non-local correlation to the adsorption energy is thus 
\bn
\Delta^\nl E^\nl_c
= E_{c,\system}^\nl -E_{c,\adenine}^\nl -E_{c,\graphite}^\nl\,,
\ee
with obvious definitions of terms.

The extension from a single adsorbant to the full adenine overlayer is
straightforward in the nonlocal calculations. For each data point we perform a total 
of five reference calculations, one for each of the
adenine molecules (because each of them has a different position relative to the 
grid) and one for the graphite layer; all reference calculations have the same unit
cell size as the main calculation. The energy contribution $\Delta^\nl E_c^\nl$ is again 
per adenine molecule. 

\subsection{Representation of an infinite sheet with an adsorbed single molecule}

The sc-GGA calculations use periodic boundary conditions whereas our implementation 
of (\ref{eq:gg}) is nonperiod. Thus in the sc-GGA calculations of $E^0$ and the 
charge density $n$ the graphite sheet is represented by a (periodic) infinite sheet.  
The representation of a single molecule on an infinite sheet requires that no 
significant inter-sheet or inter-adsorbate interactions between the supercells exist. 
The choice of a unit cell of 18 {\AA} in the direction perpendicular to the plane 
secures that within the sc-GGA calculation the graphite sheet does not interact 
with the periodic images of the system in that direction, while a supercell of 
$7\times 7$ graphite sheet unit cells (containing a total of 98 graphite sheet 
carbon atoms) in the plane of the sheet makes the inter-adsorbate interactions 
negligible (less than 0.5 meV) even for the evaluation of the non-local correlation.
With this unit cell, the minimum distance between any two atoms on two different 
adenine molecules is larger than 10 {\AA}.  

In the evaluation of $E_c^\nl$ from (\ref{eq:gg}) the electron density from
several neighboring supercells within 
the plane may be included, in order to capture the full extent of the adenine-graphite sheet 
interaction. 
Based on the decay of vdW forces at large separations, we can efficiently 
evaluate $E_c^\nl$ by 
introducing two radius cutoffs $|\br-\br'|<R_i$. Around a given point in space, 
a full grid sampling is used for the $E_c^\nl$ evaluation in the volume 
within the smallest radius, while the volume outside the smallest, but within 
the largest radius, is evaluated using a sampling of half the grid points in all 
directions. Use of $R_1 = 6.0$ {\AA} and $R_2 = 23.0$ {\AA}  
converges the contribution to the binding energy to sub-meV.
Details of the implementation is given in Ref.~\cite{berlandX}.

\section{Other computational methods}

To describe the adsorption of adenine on 
graphite, an organic molecule interacting with a chemically inert surface,
it is imperative that the vdW forces are well described. 
The vdW-DF is a first-principles DFT method,  
relieving some of the short-comings of previous \mbox{(semi-)local} approximations of the
exchange correlation term $E_{xc}$ in DFT, such as the GGA approximation.
It combines the excellent description of short-range interactions already present in GGA with 
good descriptions of the longer-ranged vdW interactions (including systems where the binding
equilibrium configuration has a range of distances over which the vdW interaction acts).  

The vdW force originates primarily from the most loosely bound electrons,
which for molecular monomers are in states modified by chemical bonding.
It is not directed through nuclear centers, as assumed by some 
semi-empirical methods. In the vdW-DF method the vdW interaction is correctly 
described as an effect originating in  
the tails of the electron distribution, and it is well suited to include 
effects of image planes \cite{KB:Molcrys}. 

In recent years the system of adenine on graphite has also been studied by 
other methods. 
The authors of Refs.\ \onlinecite{ortmann} and \onlinecite{antony} 
use semi-empirical methods which add an empirical term for the dispersion to the results of 
standard GGA-based DFT calculations.
Methods similar to that have been widely used 
\cite{brooks,halgren,scoles,kaxiras,weitao,hasegawa,zimmerli,grimme2004,TS,grimmeDFT-D3},
apparently first in 1952 \cite{brooks}.

The empirical term for the dispersion in those methods incorrectly 
assumes that the vdW interaction arises in the atomic centers.
Similarly, the often assumed notion that such forces have strengths given 
by their asymptotic free-atom forms is also undocumented. 

The earlier users of such methods recognize their
\textit{ad hoc\/} or semi-empirical nature. For example, in describing
the cutoff leading to the notion of damping functions,
Brooks \cite{brooks} writes, ``This procedure cannot be
rigorously justified, although it is certainly more reasonable than
the use of [the vdW potential] where it is divergent.''
Recent damping functions
\cite{halgren,scoles,kaxiras,weitao,hasegawa,zimmerli,grimme2004,TS,grimmeDFT-D3}
take varying forms, but most require a semi-empirical parameter for
each atom-atom pair. 

Ref.\ \onlinecite{gowtham2007} is another recent study addressing the 
adsorption of nucleobases on graphite. They use Hartree-Fock (HF) calculations
coupled with M{\o}ller-Plesset perturbation theory (MP2), in addition to LDA-based%
\footnote{LDA-based DFT cannot be used as a substitute for 
the inclusion of vdW interactions. 
As pointed out by Harris already in 1985 \cite{harris} ``LDA predicts
attraction between all systems at large separation mainly because it assigns
an unphysically long range to exchange interactions and not because, in any
sense, it simulates van der Waals interactions." This issue is summarized and discussed 
also in Ref.~\onlinecite{LangrethX}.
In some flat systems the numerical results predicted by LDA happen
to end up in the range of the physical results 
for these unphysical reasons, whereas in other geometries
the LDA gives results
that are not in accordance with experiment nor with more accurate methods 
\protect\cite{vdW0,Jesper:vdW,miao}.} DFT. 
The approach of HF with MP2 is obviously accurate if the (in principle infinite) 
graphite sheet is represented by a sufficiently large flake of graphite, but 
the approach is then also very expensive. 
Thus, to keep the computational expense down often rather small
graphite flakes are used.
The graphite flakes, terminated
by hydrogen atoms, are polycyclic aromatic hydrocarbon (PAH) molecules. 
In the following section we discuss the effect of mimicing the graphite substrate
by PAH molecules of insufficient extension. 

Results of the above studies are mentioned in and compared to our vdW-DF
results in section IV. 

\section{Results and discussions}

\begin{table}[tb]
\begin{ruledtabular}
\caption{Single adenine molecule adsorption, the values of the energy terms and distance $d$ at 
the optimal vdW-DF binding position, the binding energy $E_b$, and the
minima of the revPBE and PBE GGA calculations.
\label{tab:isolated}}
\begin{tabular}{lrr}
 Term  &   $d$ (\AA)  & $E$ (meV) \\
\hline
$\Delto E^0$ &   & 420 \\
$\Delta^\nl E^\nl_c$ &   & $-1131$\\
$E=\Delto E^0+\Delta^\nl E^\nl_c$ & 3.5 & $-711$\\
\multicolumn{3}{c}{$E_b$(single molecule) $=711$ meV}\\
\hline
$\Delto E^\GGA_\tot$ with revPBE GGA  & 5.0 &  $-11$ \\
$\Delto E^\GGA_\tot$ with PBE GGA  & 4.0 & $-45$
\end{tabular}
\end{ruledtabular}
\end{table}

The optimal molecular configuration of adenine adsorbate is determined 
by first placing the molecule relative to the graphite sheet in a 
configuration that resembles AB stacking of graphite.  
Next, we calculate the optimal distance to the graphite layer. Then we 
rotate and translate the molecule in the plane until optimal in-plane 
positions are found, within the accuracy of the method.  We here only 
consider positions with the adenine molecule parallel to the graphite plane.

\begin{figure}[tb]
\centering
\includegraphics[width=8cm]{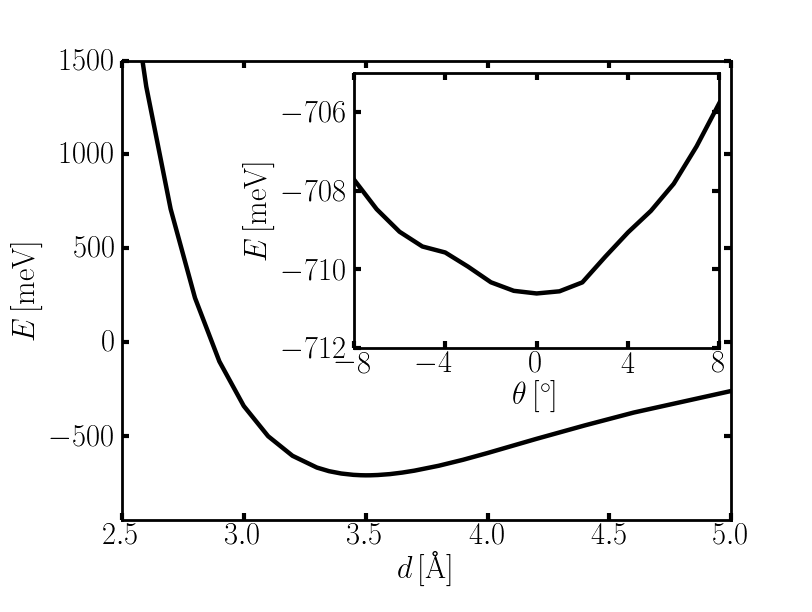}\\
\includegraphics[width=8cm]{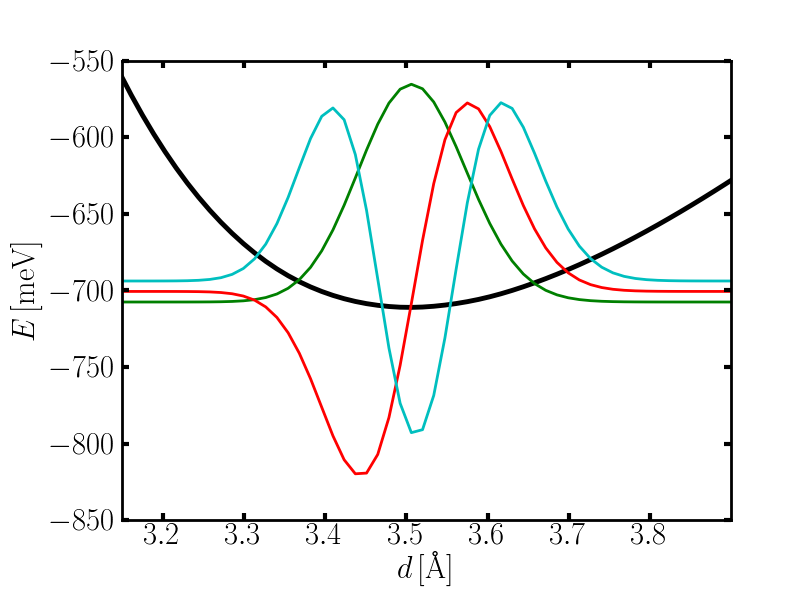}
\caption{(Color online) Top panel: Cohesive energy of adenine above a 
graphite surface at distance $d$.
The insert shows the energy variation for small in-plane rotations of the molecule around the
center of the six-fold ring close to its optimal adsorption structure. The point $0^\circ$
corresponds to the configuration used for the cohesive energy curve.
Bottom panel: Illustration of the wavefunctions of the three lowest 
vibrational states (arbitrary units on vertical axis for the wavefunctions) 
in the cohesive energy potential (black line). The wavefunctions are
 offset from the potential well bottom by their vibrational energies. 
}
\label{fig:energy}
\end{figure}

\subsection{Single molecule adsorption}
Using the updated in-plane configuration, shown in Fig.~\ref{fig:isolated}, 
we determine 
the cohesive energy curve $E(d)$ as a function of the distance to the surface in the 
direction perpendicular to the surface, $d$ (Fig.\ \ref{fig:energy}). 
We find that the molecule binds at  
$d=3.5$ {\AA} above the graphite layer, with a binding energy  $E_b= 711$ meV. 
The exchange part of revPBE is overly repulsive \cite{vdW1} at this distance and thus we expect 
our value for the binding energy to be somewhat too small. 
For naphthalene, 
an aromatic molecule with approximately the same number of electrons as adenine, 
Ref.~\onlinecite{Svetla:BenzeneGraphite} reports a binding energy of $763$ meV, using the
same vdW-DF method and choice of exchange functional as used here. 
In contrast, as shown in Table \ref{tab:isolated} for adenine and in 
Ref.~\onlinecite{Svetla:BenzeneGraphite} for naphthalene, pure GGA functionals 
such as revPBE or PBE bind at unphysically large
binding distances (4--5 {\AA}) at unphysically low binding energies ($< 50$ meV).

For the single adenine molecule, rotation around the hexagon shows little 
variation in energy (insert of Fig.~\ref{fig:energy}). 
Roughly 90\% of this small variation originates from the $E_0$ part of the total energy.
This confirms that the directional dependence of the vdW interaction is small. 

The vertical vibrational states of the adenine molecule adsorbed on the 
graphite sheet may be estimated by solving the one-dimensional Schr\"odinger 
equation for the cohesive energy potential shown in Fig.~\ref{fig:energy}. 
We find that the ground state energy when including zero-point vibrations 
is $-707.6$ meV (up from the result $-711$ meV without zero-point vibrations) 
and the first and second excited states are found at $-700.7$ meV and $-693.9$ meV.
This spectrum of lowest lying states closely resembles that of 
a harmonic oscillator at frequency around $6.8-6.9$ meV. 
 
The wavefunctions of the three lowest vibrational states are
illustrated in the bottom panel of Fig.~\ref{fig:energy}, offset with their vibrational energies.
The spatial extension of these lowest vibrational state wavefunctions is about 0.3 {\AA}.
In combination with the small energy changes for
lateral motion (illustrated also by the effect of rotational displacement of adenine, shown in
the insert of  Fig.~\ref{fig:energy}) we conclude that the precise position of the adenine
molecule on the surface of graphite has very little bearing on the binding energy.  

In this work and most of the work cited here only the interaction from one graphite layer is 
included.
If the molecules adsorb at a (multilayer) graphite surface the layers below
the top graphite layer also contribute to the interaction, but has previously been
shown to be at a very low level (3\%, as discussed in Ref.\ \onlinecite{Svetla:BenzeneGraphite}).
We therefore ignore multilayer effects here.

Other groups have studied the adsorption of single molecule on graphite using other theory methods.
Using DFT-D methods, Ortmann et al.\ \cite{ortmann} found a binding energy of 1.01 eV,  
and a separation of 3.4 {\AA} while more recently Antony et al.\ \cite{antony} found binding
at 0.91 eV and 3.0 {\AA}, both results are for single molecules adsorbed on a sheet of graphite. 
Unlike the above-mentioned and the present study, the MP2 calculations of Gowtham et al.\ \cite{gowtham2007} 
use an adenine molecule with a methyl group attached. They find a binding 
energy 0.94 eV and separation 3.5 {\AA}. 

\subsection{Graphite size convergence test study} 

\begin{figure}
\begin{centering}
\includegraphics[width=8cm]{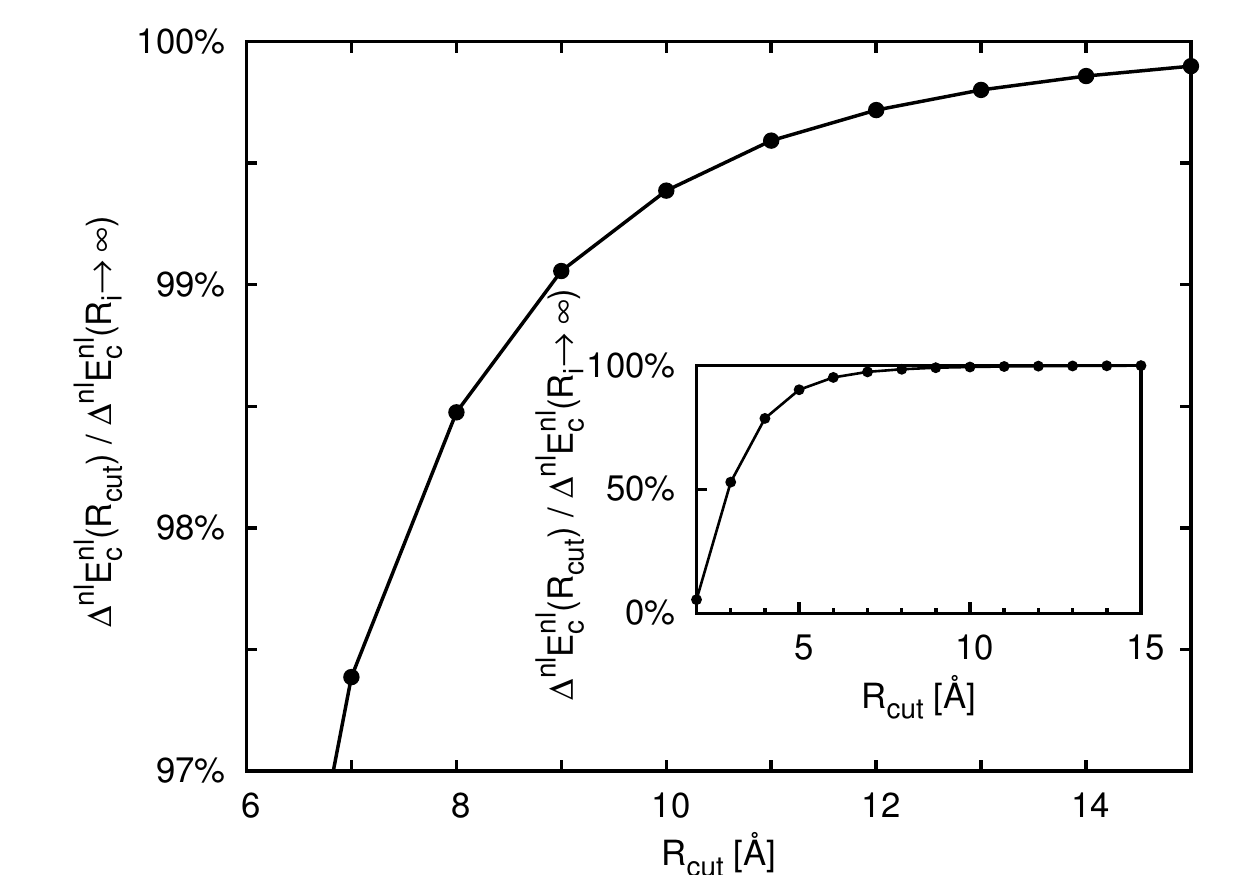}
\caption{Convergence of $\Delta^\nl E_c^\nl$ with graphite sheet cutoff radius
$R_\cut$,  relative to a calculation with converged values of the cutoff radii, 
 $R_1$ and $R_2$. The use of converged values 
of $R_1$ and $R_2$ is symbolically denoted ``$R_i\rightarrow\infty$".
The calculations are carried out for a single adenine molecule on a sheet of graphite
at the optimal position ($d= 3.5$ {\AA}). \label{fig:conv}}
\end{centering}
\end{figure}

We use calculations of the long-range correlation contribution, 
$\Delta^\nl E_c^\nl$, to estimate the effect of using small 
PAH molecules as substitutes for the graphite sheet in 
the MP2 calculations of Ref.~\onlinecite{gowtham2007} and other similar studies. 
We perform a number of crude but generous test calculations.
The tests are available directly from the vdW-DF method simply by restricting the 
cutoff radii $R_1$ and $R_2$ in the $\Delta^\nl E_c^\nl$ calculations by the value $R_\cut$. 
 
Figure \ref{fig:conv} illustrates the convergence of $\Delta^\nl E_c^\nl$
with $R_\cut$. The value of $R_\cut$ sets 
the amount of interactions with the graphite sheet included.
For $R_\cut \approx 9$~{\AA} Fig.~\ref{fig:conv} shows that about 1\% of the interaction  
is discarded. 
In the production runs we find that at more than 6 {\AA} distance
between interacting points only every second grid point in each direction 
needs to be included, which is reflected in our choice of $R_1=6$ {\AA}
(and $R_2=23$~{\AA}). For our convergence tests we further do not include
any pairs of points in space $\mathbf{r}$ and $\mathbf{r}'$
that have $|\mathbf{r}-\mathbf{r}'|>R_\cut$. The value $R\cut=9$~{\AA} 
corresponds approximately to the use of a 96 C atom PAH molecule as a representant
for the graphite sheet, for which we therefore predict that $\sim 1$\% of the interaction
(compared to graphite) is lost.  

For smaller $R_\cut$ the convergence is much worse, as illustrated in
the insert of Figure \ref{fig:conv}.   
If we include only a radius of 4 {\AA}, corresponding approximately to the 
inclusion of 24--30 C atoms of the graphite sheet in the $\Delta^\nl E_c^\nl$ calculation (\ref{eq:gg}) 
we loose 21\% of the interaction contribution. We therefore estimate that the use of a 
28 C atom PAH molecule in 
Ref.~\onlinecite{gowtham2007} misses an important part of the
long-range interaction compared to the use of graphite or a larger size of PAH molecule.  

Our convergence tests are not fully equivalent to using PAH molecules to model the graphite
layer, such as done in the MP2 calculations. 
This is because for this $E_c^\nl$ calculation  
any point on
the adenine molecule is paired with all grid points at a distance less than $R_\cut$,
even those outside the volume covered by a PAH model substitute.
At a specified cutoff radius $R_\cut$ this test is therefore more generous than a similar
calculation using a PAH molecule having roughly the radius $R_\cut$.

Another test of the effect of using small PAH molecules instead of graphite
as a substrate is reported in Ref.~\onlinecite{antony}, using actual PAH molecules but 
an empirical dispersion term (via DFT-D). There, similar results were found: 
reducing the size of the PAH molecule from 150 C atoms to 24 C atoms 
caused a loss of 24\% of the dispersion interaction (from $-32.9$ kcal/mol to $-24.9$ kcal/mol 
for adenine).
The use of a 150 C atom PAH molecule roughly corresponds to the value 11 {\AA} of our cutoff 
radius $R_\cut$, where we find minimal ($\sim 0.4$\%) loss of interaction compared to 
a converged size (Fig.~\ref{fig:conv}).

\subsection{Adenine overlayer}

Some molecular adsorbates, such as adenine molecules \cite{tao1994,freund}, can 
spontaneously form an ordered lattice on a graphite surface. Here we present 
an estimate of the formation energy of the two-dimensional adenine crystal. This calculation 
illustrates the potential of the vdW-DF method to discern the different phases of adsorbate 
crystals and can therefore lend credibility to interpretations of 
scanning-tunneling microscopy images.

Figure \ref{fig:crystal} shows the molecular position on the surface of 
graphite for the two-dimensional adsorbate crystal. The crystal symmetry is 
chosen to be the same 
as that determined with the force-field calculations in Ref.~\onlinecite{freund}, 
but with the molecules each in a configuration relative to the graphite layer identical 
to our single adsorbate result.

\begin{table}[tbh]
\begin{ruledtabular}
\caption{Various contributions to the adenine overlayer binding energy $E_b$, per adenine 
molecule. Values at the optimal graphite-overlayer separation $d=3.5$ {\AA}.
\label{tab:overlayer}}
\begin{tabular}{lr}
Term & Energy (meV)\\
\hline
$\Delta^\nl E^\nl_c$  & $-1780$ \\
$\Delto E^0_\overlayersheet$   & 482 \\
$\Delto E^0_\sheetchain$   & 199 \\
$\Delto E^0_\chainmolecule$   & 93  \\
$\Delto E^0 = \Delto E^0_\overlayersheet +\Delto E^0_\sheetchain+\Delto E^0_\chainmolecule $  & 774\\
$E=\Delto E^0+\Delta^\nl E^\nl_c$ & $-1006$\\
\multicolumn{2}{c}{$E_b$(overlayer) $= 1006$ meV} \\
\hline
$E_b$(overlayer) - $E_b$(single molecule)  & 295 \\
$E_b$(free-floating crystal) & 239 \\
 \end{tabular}
\end{ruledtabular}
\end{table}

In the process of calculating the cohesive energy $E$ (the difference between the adenine overlayer on 
the graphite sheet and the adenine molecules all lifted off individually) a number of partial energy terms 
are calculated, corresponding to the terms illustrated in Fig.\ \ref{fig:assembly}. These 
partial energy terms are provided in Table \ref{tab:overlayer}. 

We determine the binding energy of the adenine overlayer to be 1006 meV 
per molecule, which is larger than
that of an isolated adenine molecule. 
The energetic gain of the system when single, adsorbed molecules 
are moved together to form an overlayer crystal (the overlayer formation energy), 
is found to be 295 meV 
per molecule in our not-fully optimized overlayer crystal structure.
This result shows that a two-dimensional ordered crystal structure is 
energetically much more preferred than isolated molecules
on the surface, in agreement with experimental findings showing spontaneous formation of the 
adenine crystal overlayers \cite{tao1994,freund}. Using TDS the 
binding energy of adenine clusters on graphite 
has been measured to be 23.2 kcal/mol (1006 meV per molecule) \cite{freundthesis}, 
in very good agreement with our results for the crystal overlayer.

The value of the overlayer formation energy may be compared to the 
energy gained by creating a free-floating
two-dimensional adenine crystal (with the same structure as the overlayer) from isolated molecules, 
$E_b$(free-floating crystal) $= 239$ meV/molecule. The gain of assembling the crystal \textit{on\/} the 
graphite surface (295 meV/molecule) instead of away from the graphite (239 meV/molecule) 
is a mere 56 meV/molecule, not insignificant 
but clearly smaller than the effect of the mutual binding of the adenine molecules. Of course, if the 
free-floating adenine molecules were allowed to assemble in the most optimal structure, the molecules 
would stack and the gain in binding energy would increase to about 300 meV/molecule, depending on 
the details \cite{jacs}. 

\section{Conclusions}

We use the first-principles vdW-DF method to study the adsorption of adenine on graphite.
We find that the adenine molecule is physisorped at a distance 3.5 {\AA} above the graphite surface.
We also find that the physisorption well is shallow and therefore small changes in position (all 
directions and rotations) lead to only small changes in adsorption energy; the 
molecule is mobile.

Our calculations show an adsorption energy of 711 meV per molecule for adenine molecules far apart on the
surface, whereas molecules forming a two-dimensional overlayer cluster gain significantly more: 
1006 meV per molecule, both situations compared to molecules floating off as a dilute gas. 

The small barriers for changing the position, mentioned above, 
along with this 295 meV/molecule gain per molecule
for moving molecules closer together, is consistent with the tendency of
adenine on graphite to assemble into clusters of two-dimensional overlayers. 
We find that although it is more favorable for the adenine molecules to form the
overlayer crystal at the graphite surface, the largest part of the energy gain,
about 239 meV/molecule,  
is also obtained in the (unphysical) situation of  the adenine molecules being moved together  
into the same positions but without having a graphite surface nearby. The role of the graphite 
surface in forming clusters therefore seems to be mainly to attract the molecules and orient them 
(flat on the surface) before assembly into cluster,
rather than contributing any major part to the cluster formation energy. 
    
In a crude estimate of the effect of using small PAH molecules to model the graphite surface we found
that an important part of the long-ranged correlation effects are lost in such models. If a PAH molecule is
used to model graphite, it must be significantly larger than the adsorbed molecule: in the case of adenine we estimate that 
a 96 C atom PAH molecule is the smallest acceptable, and for full convergence an even larger PAH molecule
should be used. For comparison, a 96 C atom PAH molecule has the approximate radius 9 {\AA}
and the adenine radius is approximately 2.5 {\AA}. 

In summary, we find adsorption energies of adenine on graphite using vdW-DF. The energies are calculated
both for single adenine molecules on a graphite sheet and a two-dimensional crystal overlayer of adenine
on a graphite sheet. The adsorption energy is highest per molecule for the overlayer compared to single
molecular adsorption, in agreement with the tendency to cluster formation seen in experiment.

\acknowledgments
We thank P.\ Hyldgaard and B.I.\ Lundqvist for useful discussions. Partial support from the 
 Swedish Research
Council (VR) to ES and SC is gratefully acknowledged. We also acknowledge
the allocation of computer time at UNICC/C3SE (Chalmers) and SNIC (Swedish National
Infrastructure for Computing) and funding from SNIC for KB's participation in the 
national graduate school NGSSC. 
Work at Rutgers supported by NSF Grant DMR-0801343.

\end{document}